\documentclass[a4paper,11pt]{article}
\pdfoutput=1

\usepackage{jcappub}
\usepackage[T1]{fontenc}

\newcommand{\pchip}{\texttt{PCHIP}}

\begin{document}
\title{Was there an early reionization component in our universe?}

\author[a]{Pablo Villanueva-Domingo}
\author[a]{Stefano Gariazzo}
\author[b,c,d]{Nickolay Y.\ Gnedin}
\author[a]{Olga Mena} 
\affiliation[a]{Instituto de F\'isica Corpuscular (IFIC), CSIC-Universitat de Val\`encia,\\
Apartado de Correos 22085,  E-46071, Spain}
\affiliation[b]{Particle Astrophysics Center, Fermi National Accelerator Laboratory,\\ Batavia, IL 60510, USA}
\affiliation[c]{Kavli Institute for Cosmological Physics, The University of Chicago,\\ Chicago, IL 60637, USA}
\affiliation[d]{Department of Astronomy \& Astrophysics, The University of Chicago,\\ Chicago, IL 60637, USA} 

\emailAdd{pablo.villanueva@ific.uv.es}
\emailAdd{gariazzo@ific.uv.es}
\emailAdd{gnedin@fnal.gov}
\emailAdd{omena@ific.uv.es}

\abstract{A deep understanding of the Epoch of Reionization is still missing in our knowledge of the universe.
While future probes will allow us to test the precise evolution of the free electron fraction from redshifts between $z\simeq 6$ and $z\simeq 20$, at present one could ask what kind of reionization processes are allowed by present Cosmic Microwave Background temperature and polarization measurements.
An early contribution to reionization could imply a departure from the standard picture where star formation determines the reionization onset. 
By considering a broad class of possible reionization parameterizations, we find that current data do not require an early reionization component in our universe and that only one marginal class of models, based on a particular realization of reionization, may point to that.
In addition, the frequentist Akaike Information Criterion (AIC) provides strong evidence against alternative reionization histories, favoring the most simple reionization scenario, which describes reionization by means of only one (constant) reionization optical depth $\tau$.}

\maketitle

\section{Introduction}
\label{sec:intro}
The Epoch of Reionization is the interval of time during which the cosmic gas evolves from an almost completely neutral state (neglecting the recombination leftovers) to an ionized state.
This ionization process is believed to happen due to the onset of star formation at redshifts $z\simeq 12$, and it is believed to last until $z\simeq 6$.
Several astrophysical observables (quasars~\cite{Fan:2005es,Becker:2014oga}, Lyman $\alpha$ emitters~\cite{Stark:2010qj,Treu:2013ida,Pentericci:2014nia,Schenker:2014tda,Tilvi:2014oia}, $\gamma$ ray bursts~\cite{Wang:2015ira,Gallerani:2009aw}) seem to agree with this hypothesis.
However, the precise details of the overall reionization process still remain obscure.
The main reason is that the currently available most precise information on the reionization period comes from Cosmic Microwave Background (CMB) measurements through a redshift-integrated quantity.
During reionization, the number density of free electrons which can scatter the CMB, $n_e$, increases. As a consequence, the reionization optical depth $\tau$ increases according to a line of-sight integral of $n_e$, generating a suppression of the CMB peaks at any scale within the horizon at the reionization period.
This suppression, however, can be easily compensated with an enhancement of the primordial power spectrum amplitude, $A_{\rm s}$.
A much better and cleaner measurement of $\tau$ can be obtained via measurements of the CMB polarization, which is linearly affected by reionization (see e.g. Refs.~\cite{Kaplinghat:2002vt,Haiman:2003ea,Holder:2003eb,Hu:2003gh} for seminal works and \cite{Reichardt:2015cos} for a recent review). The latest measurements of the Planck collaboration provide a value of $\tau = 0.055 \pm 0.009$~\cite{Aghanim:2016yuo, Adam:2016hgk} based exclusively on the CMB polarization spectrum.
This value of $\tau$ is in a much better agreement than previous WMAP~\cite{Hinshaw:2012aka} and Planck~\cite{Ade:2015xua} estimates with observations of Lyman-$\alpha$ (Ly-$\alpha$) emitters at $z\simeq 7$~\cite{Stark:2010qj,Treu:2013ida,Pentericci:2014nia,Schenker:2014tda,Tilvi:2014oia}, which require that reionization is complete by $z\simeq 6$.
Even if now cosmological and astrophysical tests of the reionization process seem to agree, the measurement of $\tau$ provides only integrated information on the free electron fraction $x_e$, and not on its precise redshift evolution.
Consequently, the same measured value of $\tau$ may correspond to very different reionization histories.

Traditionally, the most commonly exploited model for the time evolution of the free electron fraction, $x_e(z)$, uses a step-like transition, implemented via a hyperbolic tangent~\cite{Lewis:2008wr}.
Model independent attempts have been carried out in several works in the past~\cite{Hu:2003gh,Mortonson:2007hq,Mortonson:2007tb,Mortonson:2008rx,Mortonson:2009qv,Mortonson:2009xk,Mitra:2010sr,Lewis:2006ym,Pandolfi:2010dz,Pandolfi:2010mv} and also more recently~\cite{Heinrich:2016ojb,Hazra:2017gtx,Mitra:2017oxx}, based either on a redshift-node decomposition of $x_e(z)$ or on a Principal Component Analysis (PCA) of the CMB polarization angular power spectrum.
More concretely, using the latter approach, the authors of \cite{Heinrich:2016ojb} claimed that Planck 2015 data favors a high-redshift ($z>15$) component to the reionization optical depth.
The quoted $2\sigma$ evidence would come from the excess in power in the low multipole range of the Planck 2015 CMB polarization spectrum. 
Accordingly to their results, the functional form of the usual step-like model prevents a priori for such an early component in the reionization history of our universe.
However, the authors of \cite{Hazra:2017gtx}, using a different method, which implements reionization through a non-parametric reconstruction that uses a Piecewise Cubic Hermite Interpolating Polynomial (\pchip), find only marginal evidence for extended reionization histories.
Since an early component in the reionization history $x_e(z)$ (or, in other words, a high redshift contribution to the reionization optical depth $\tau$) may either imply the need for a high-redshift population of ionizing sources (hypothesis that will be tested by the future James Webb Space Telescope~\cite{Gardner:2006ky}),
or give hints about a possible energy injection from dark matter annihilations or decays~\cite{Pierpaoli:2003rz,Mapelli:2006ej,Natarajan:2008pk,Natarajan:2009bm,Belikov:2009qx,Huetsi:2009ex,Cirelli:2009bb,Kanzaki:2009hf,Natarajan:2010dc,Giesen:2012rp,Diamanti:2013bia,Lopez-Honorez:2013lcm,Lopez-Honorez:2016sur,Poulin:2016nat,Poulin:2015pna},
or accreting massive primordial black holes~\cite{Ricotti:2007au,Horowitz:2016lib,Ali-Haimoud:2016mbv,Blum:2016cjs,Poulin:2017bwe},
it is mandatory to robustly establish what current data prefer, regardless of the model used to describe the redshift evolution of the free electron fraction. 

Here we first analyze several possible parameterizations for reionization (PCA with several fiducial cosmologies and the \pchip\ method)
and explore the corresponding constraints on the reionization history of the universe.
We then shall exploit tools related to model selection among competing models, using both the Akaike Information Criterion (AIC) and the Bayesian Information Criterion (BIC), which will allow us to quantitatively decide which model is currently preferred and whether it exists or not an indication for an early reionization component in our universe.

The structure of the paper is as follows.
We start by discussing the different reionization approaches that we shall test against current data in Sec.~\ref{sec:histories}.
In Sec.~\ref{sec:data} we describe the cosmological observations exploited in our numerical analyses, whose results are shown in Sec.~\ref{sec:results}.
Our conclusions are summarized in Sec.~\ref{sec:conclusions}.

\section{Reionization histories}
\label{sec:histories}

In the following, we will derive the constraints on the reionization history of our universe from cosmological observations exploring several possible scenarios, focusing on a possible early reionization component in our universe.
For that, we shall exploit the reionization optical depth:
\begin{equation}
\tau(z) = \int_z^{\infty} dz' \frac{c ~dt'}{dz'}  (n_{\rm e}(z')- n_{\rm e, 0}(z'))\sigma_{\rm T}\,~,
\label{eq:cumtau}
\end{equation}
where $n_{\rm e}(z)=n_{\rm H}(0)(1+z)^3x_{\rm e}(z)$ and $n_{\rm e,0}(z)=n_{\rm H}(0)(1+z)^3x_{\rm e, 0}(z)$, being $n_{\rm H}(0)$ the number density of hydrogen at present, $x_{\rm e}(z)$ the free electron fraction and $x_{e,0}(z)$ the free electron fraction leftover from the recombination epoch (see e.g.\ \cite{Kolb:1990vq,2009fflr,2010gfe}). Therefore, Eq.~\eqref{eq:cumtau} just accounts for the cumulative Compton optical depth after recombination, subtracting the pre-reionization contribution.

\subsection{Canonical scenarios}
\label{subsec:canonical}
We start describing the free electron fraction by means of the most simple and commonly exploited parameterizations in the literature, i.e.\ the so-called \emph{redshift-symmetric} and \emph{redshift-asymmetric} parameterizations (see e.g.~\cite{Adam:2016hgk}).   

\begin{itemize}

\item \emph{Redshift-symmetric} parameterization.

The most economical and widely employed approach to describe the reionization process in our universe assumes that the free electron fraction follows a step-like function, taking the recombination leftover value at high redshifts and becoming close to one at low redshifts, and being described by the hyperbolic tangent function~\cite{Lewis:2008wr}
\begin{equation}
x_e^{\rm tanh}(z) = \frac{1+f_{\rm He}}{2} \left(1+ \tanh \left[ \frac{y(z_{\rm{re}})-y(z)}{\Delta y} \right] \right),
\label{eqn:tanh}
\end{equation}
where $y(z)=(1+z)^{3/2}$, $\Delta y=3/2(1+z_{\rm{re}})^{1/2}\Delta z$, and $\Delta z$ is the width of the transition, fixed in the following to $\Delta z=0.5$.
This parametrization is named ``redshift symmetric'' because the redshift interval between the beginning of reionization and its half completion equals the corresponding one between half completion and the reionization offset, and it is the default one implemented in Boltzmann solver codes such as \texttt{CAMB}~\footnote{\href{http://camb.info}{http://camb.info}}~\cite{Lewis:1999bs}.
This parameterization, as well as the following ones, also accounts for the first ionization of helium $f_{\rm He}=n_{\rm{He}}/n_{\rm{H}}$, assumed to happen at the same time than that of hydrogen.
The full helium reionization is modeled via another hyperbolic tangent function with $z_{\rm{re,He}}=3.5$ and $\Delta z=0.5$.
Therefore, the only free parameter in this simple approach is the reionization redshift $z_{\rm{re}}$.
When this redshift-symmetric parameterization is used as the fiducial model in our PCA analyses (see next subsection), we fix $z_{\rm{re}}=8.8$, following the results quoted in Ref.~\cite{Adam:2016hgk}.

\item \emph{Redshift-asymmetric} reionization.

Besides the previous case, alternative reionization parameterizations with a non redshift-symmetric transition have been proposed in the literature.
One of the most flexible choices, which shows good agreement with current measurements from quasars, Ly$\alpha$ emitters and star-forming galaxies, is represented by a power law, described via three parameters~\cite{Adam:2016hgk,Douspis:2015nca}:
\begin{equation}
  x_e^{asym}(z) =
  \begin{cases}
	1+f_{\rm He} & \mbox{for } z<z_{\rm end}, \\
    (1+f_{\rm He}) \left(\frac{z_{\rm early}-z}{z_{\rm early}-z_{\rm end}}\right)^\alpha
      & \mbox{for } z_{\rm end} < z < z_{early}, \\
      0 & \mbox{for } z > z_{\rm early}.
  \end{cases}
  \label{eqn:asym}
\end{equation}
Following Planck 2016 reionization analyses~\cite{Adam:2016hgk}, when using this redshift-asymmetric model as a fiducial model in our PCA analyses, we shall fix the redshift at which the first sources in our universe switch on, $z_{\rm early} = 20$, the redshift at which reionization is fully complete, $z_{\rm end} = 6$, and the exponent $\alpha = 6.10$. 
\end{itemize}

\subsection{Principal Component Analysis (PCA)}
The second method we follow here to model the reionization process is the Principal Component Analysis (PCA) approach of Refs.~\cite{Hu:2003gh,Mortonson:2007hq,Mortonson:2007tb,Mortonson:2008rx,Mortonson:2009qv,Mortonson:2009xk,Mitra:2010sr}, exploited more recently in Refs.~\cite{Heinrich:2016ojb,Mitra:2017oxx}.
Following these previous works, we discretize the redshift range from $z_{\rm{min}}=6$ to $z_{\rm{max}}=30$ in $N_z$ bins of width of $\delta z = 0.25$.
We set the ionization fraction to $x_e=0$ for $z \geq z_{\rm{max}}$, when the reionization processes have not started yet, while for $z \leq 6$ we assume fully ionized hydrogen and singly ionized helium, i.e.\ $x_e=1+f_{\rm He}$.
The full helium reionization is modeled as aforementioned.
This approach makes use of the Fisher information matrix~\cite{Tegmark:1996bz}, that we compute as:
\begin{equation}
F_{ij} = \sum_{\ell=2}^{\ell_{\rm max}}\frac{1}{\sigma_{ C_{\ell}}^2}
 \frac{\partial  C_{\ell}}{\partial x_e(z_i)}
 \frac{\partial  C_{\ell}}{\partial x_e(z_j)} =\sum_{\ell=2}^{\ell_{\rm max}}\left(\ell+\frac{1}{2}\right)
 \frac{\partial \ln C_{\ell}}{\partial x_e(z_i)}
 \frac{\partial \ln C_{\ell}}{\partial x_e(z_j)} \,,
\label{eq:fisher}
\end{equation}
where the $C_{\ell}$ are the components of the large angle $EE$ polarization spectrum.
The sum above is truncated at $\ell_{\rm max}=100$, because the reionization imprint is mostly located in the lowest modes of the CMB polarization spectrum.
In Eq.~\eqref{eq:fisher} we have used the well-known result for the cosmic variance: $\sigma_{ C_{\ell}}^2 = C_{\ell}^2\, 2/(2\ell+1)$.
Having the Fisher matrix, we can diagonalize it and find that the eigenfunctions are the principal components $S_{\mu}(z)$ and the eigenvalues are proportional to the inverse of the estimated variance of each eigenmode, $\sigma^2_{\mu}$.
Using the normalization of Ref.~\cite{Mortonson:2007hq}, we can write the Fisher matrix as
\begin{equation}
F_{ij}=\frac{1}{(N_z+1)^2}\sum_{\mu=1}^{N_z}
 \frac{1}{\sigma^2_{\mu}}S_{\mu}(z_i) S_{\mu}(z_j)~.
\label{eq:fisher2}
\end{equation}
We sort the different eigenfunctions in order to have the smallest uncertainties at the lowest modes, being therefore the $\mu=1$ case the best constrained mode.
Due to completeness and orthogonality of the principal components, the following properties are fulfilled:
\begin{align}
\int_{z_{\rm min}}^{z_{\rm max}} dz \, S_{\mu}(z)S_{\nu}(z)&=(z_{\rm max}-z_{\rm min})\delta_{\mu\nu} \, , \\
\sum_{\mu=1}^{N_z} S_{\mu}(z_i)S_{\mu}(z_j)&= (N_z+1) \delta_{ij}\,.
\end{align}
Since the width of the bins is chosen to be sufficiently small, in practice we can replace the integrals over redshift by discrete sums.
One of the ideas behind the PCA approach is that one can write redshift-dependent quantities such as the ionization fraction as a linear combination of the principal components.
Since the lowest modes have the smallest uncertainties, we truncate the sum, using only the first 5 principal components, following Ref.~\cite{Mortonson:2007hq}.
We apply the PCA analysis to the ionization history in two different ways, which are explained below.

\begin{itemize}

\item \textbf{Case A}

In the first PCA approach, named \textbf{PCA-A} in what follows, the reionization history reads as
\begin{equation}
x_e^A (z) = x^{\rm{fid}}_{\rm e} (z)+ \sum_\mu m_{\mu}^{A} S_\mu (z)~.
\label{eq:pca_a}
\end{equation}
Given a fiducial model $x^{\rm{fid}}_{\rm e} (z)$, and knowing the amplitudes derived from the Fisher matrix (see Eq.~\eqref{eq:fisher}), one can recover an arbitrary ionization history using a PCA analysis.
This is the standard approach adopted in Refs.~\cite{Mortonson:2007hq,Heinrich:2016ojb} in order to constrain the ionization history with CMB data.
Following \cite{Mortonson:2007hq}, we can derive upper and lower bounds for each amplitude $m_{\mu}$:
\begin{equation}
m_{\mu}^{\pm} = \int_{z_{\rm min}}^{z_{\rm max} } dz \frac{S_{\mu}(z)[x_e^{\rm max} -2 x_e^{\rm fid}(z)]
\pm x_e^{\rm max} | S_{\mu}(z)|}{2(z_{\rm max}-z_{\rm min})}~.
\label{eq:mbounds}
\end{equation}
Additionally, in order to guarantee physical ionization histories, the choice of our amplitudes $m_{\mu}$ has to fulfill the condition $0 \leq x_e(z) \leq 1+f_{\rm He}$ at any redshift $z$~\footnote{Notice that this constraint for physicality is stronger than that followed in Ref.~\cite{Heinrich:2016ojb}, as any unphysical model will be retained for the Monte Carlo analyses.}.

\item \textbf{Case B}

In the second of our PCA analyses, named \textbf{PCA-B}, we choose a different approach to the standard PCA analysis described above, in which the free electron fraction is proportional to the fiducial model plus the PCA decomposition.
Here, we exploit the functional form of the fiducial model in order to test other possible reionization parameterizations.
Following this idea, for the redshift-symmetric, \textit{tanh} description, we insert the PCA decomposition inside the argument of the hyperbolic tangent:
\begin{equation}
x_e^{B,tanh}(z) = \frac{1+f_{\rm He}}{2} \left(1+ \tanh \left[ \frac{y(z_{\rm{re}})-y(z)}{\Delta y} + \sum_\mu m_\mu^B S_\mu (z)  \right] \right)~.
\label{eqn:tanh_b}
\end{equation}
Notice that we recover the fiducial \textit{tanh} model by setting the amplitudes $m_{\mu}$ to $0$.
We perform an analogous replacement for the redshift-asymmetric parameterization:
\begin{equation}
  x_e^{B,asym}(z) =
  \begin{cases}
	1+f_{\rm He} & \mbox{for } z<z_{\rm end}, \\
    (1+f_{\rm He}) \left(\left(\frac{z_{\rm early}-z}{z_{\rm early}-z_{\rm end}} \right)+ \sum_\mu m_\mu^B S_\mu (z)\right)^\alpha
      & \mbox{for }z_{\rm end} < z < z_{\rm early}, \\
      0 & \mbox{for } z > z_{\rm early}.
  \end{cases}
  \label{eqn:asym_b}
\end{equation}
We take for the specific parameters of the \textit{tanh} and \textit{asym} cases the fiducial values given in Sec.~\ref{subsec:canonical}.

\end{itemize}

\subsection{\pchip}
The third and last method we adopt in order to describe the reionization history is based on a non-parametric form for the free electron fraction $x_e(z)$, which is described using the function values $x_e(z_i)$ in a number $n$ of fixed redshift points $z_1,\ \ldots,\ z_n$.
Following the procedure adopted for the PCA analyses, we fix the function to be a constant
both at low
redshifts ($z\leq6$) and at high redshifts ($z\geq30$).
The first and the last redshift nodes we use to parameterize the function at intermediate redshifts are therefore $z_1=6$ and $z_n=30$, where we also want the function to be continuous:
as a consequence, the values $x_e(z_1)=1+f_{\rm He}$ and $x_e(z_n)=0$ are fixed
and the number of varying parameters that describe $x_e(z)$ is always $n-2$.
We consider a case with a total of $n=7$ nodes (5 free parameters),
located at redshifts
\begin{equation}\label{eq:nodes7}
 z_i \in \{6,\,7,\,8.5,\,10,\,13,\,20,\,30\}\,,
\end{equation}
in order to have the same number of free parameter than in the PCA cases.

The function $x_e(z)$ at $z\neq z_i$ is computed through an interpolation among its values in the nodes.
We employ the
``piecewise cubic Hermite interpolating
polynomial'' (\pchip)~\cite{Fritsch:1980,Fritsch:1984}
in a very similar way to Refs.~\cite{Gariazzo:2014dla,DiValentino:2015zta,Gariazzo:2015qea,DiValentino:2016ikp},
where the \pchip\ function was adopted to describe the power spectrum
of initial curvature perturbations, or the more recent work of \cite{Hazra:2017gtx}, where the \pchip\ method has also been used to model the evolution of $x_e(z)$.
The idea behind the \pchip\ function is similar to that of the natural cubic spline,
with the difference that the monotonicity of the series of interpolating points
must be preserved.
Spurious oscillations that may be introduced by the standard spline
are avoided by imposing a condition on the first derivative of the function in the nodes,
which must be zero if there is a change in the monotonicity
of the point series.
A more detailed discussion on the \pchip\ function can be found in the appendix of Ref.~\cite{Gariazzo:2014dla}.

Summarizing, the free electron fraction in the \pchip\ case is described by:
\begin{equation}\label{eq:xe_pchip}
  x_e(z) =
  \begin{cases}
    1+f_{\rm He}
    & \mbox{for } z \leq z_1, \\
    \pchip(z;\ x_e(z_1),\ \ldots,\ x_e(z_n))
    & \mbox{for } z_1 < z < z_n, \\
    0
    & \mbox{for } z \geq z_n,
  \end{cases}
\end{equation}
where $n$ will be 7 and the redshifts $z_i$ are reported in Eq.~\eqref{eq:nodes7}.

For the values of the function in the varying nodes,
which are the free reionization parameters in our Markov Chain Monte Carlo analyses,
we impose a linear prior $0 \leq x_e(z_i) \leq 1+f_{\rm He}$,
with $i=2,\ \ldots,\ n-1$.
This ensures that the free electron fraction is always positive and smaller than its value today.
The value of the reionization optical depth $\tau$  
that we report in our results is derived from Eq.~\eqref{eq:cumtau}.

\section{Cosmological data}
\label{sec:data}
We use Planck satellite 2015 measurements of the CMB temperature,
polarization, and cross-correlation spectra~\cite{Adam:2015rua,Ade:2015xua}
to derive the constraints on the possible reionization histories~\footnote{%
We make use of the publicly available Planck likelihoods~\cite{Aghanim:2015xee}, see \href{http://www.cosmos.esa.int/web/planck/pla}{www.cosmos.esa.int/web/planck/pla}.
}.
More precisely, we exploit both
the high-$\ell$ ($30 \leq \ell \leq 2508$) and
the low-$\ell$ ($2 \leq \ell \leq 29$) $TT$
likelihoods
based on the reconstructed CMB maps
and
we include the Planck
polarization likelihoods in the low-multipole regime
($2 \leq \ell \leq 29$), plus the high-multipole ($30 \leq \ell \leq 1996$) $EE$ and $TE$ likelihoods~\footnote{The latest reionization constraints from the Planck collaboration do not consider the TE data in the analyses, due to its larger cosmic variance and its weaker dependence on the reionization optical depth, when compared to EE measurements, see \cite{Adam:2016hgk}.}. 
All these CMB likelihood functions depend on several nuisance parameters
(e.g.\ residual foreground contamination, calibration, and
beam-leakage~\cite{Ade:2015xua,Aghanim:2015xee}),
which have been properly considered and marginalized over.  
To derive constraints on the reionization history and related parameters, we have modified the Boltzmann equations solver \texttt{CAMB} code \cite{Lewis:1999bs} and apply
Markov Chain Monte Carlo (MCMC) methods by means of an adapted version of the \texttt{CosmoMC} package~\cite{Lewis:2002ah}.
As for current constraints, we consider a minimal version of the $\Lambda$CDM model, described by the following set of parameters: 
\begin{equation}\label{parameterPPS}
\{\omega_{\rm{b}},\,\omega_{\rm{c}},\, \Theta_{\rm{s}},\,\ln{(10^{10} A_{\rm{s}})},\,n_{\rm{s}}\}~,
\end{equation}
where $\omega_{\rm{b}}\equiv\Omega_{\rm{b}}h^2$ and $\omega_{\rm{c}}\equiv\Omega_{\rm{c}}h^2$
represent the physical baryon and cold dark matter energy densities, $\Theta_{\rm{s}}$
is the angular scale of recombination, $A_{\rm{s}}$ is the primordial power spectrum amplitude and $n_{\rm s}$ the spectral index.
Notice that we do not have $\tau$ among the parameters included in our analyses, as $\tau$ is a derived parameter.
Instead, we will add the additional parameters describing the PCA and \pchip\ reionization models, that will lead to the constraints presented in what follows.    

\section{Results}
\label{sec:results}
\begin{figure}[t]
\centering 
\includegraphics[width=0.85\textwidth]{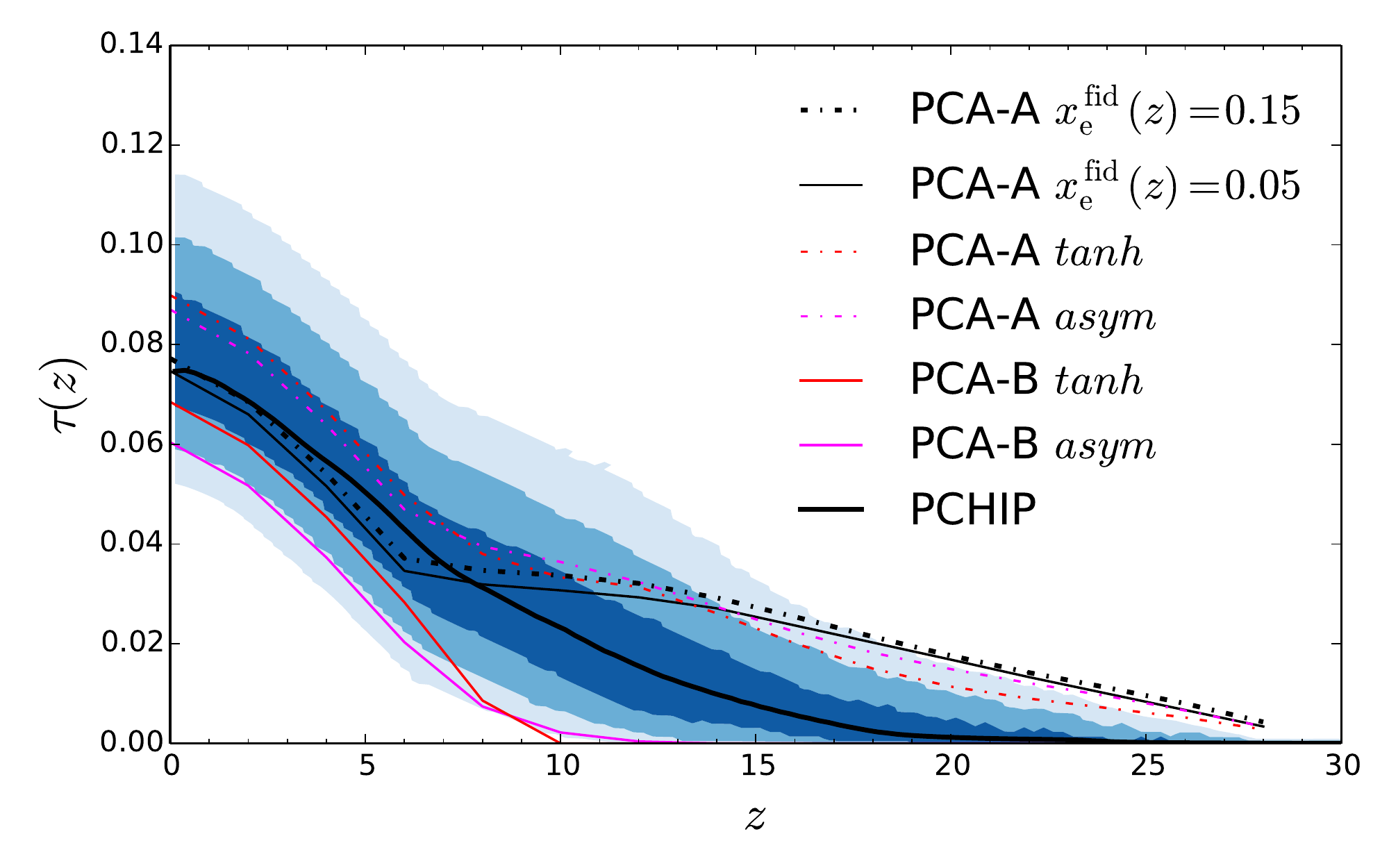}
\caption{\label{fig:tau} Cumulative redshift evolution of the reionization optical depth $\tau(z)$ for several possible reionization scenarios.
The black thin solid and dot-dashed lines illustrate the PCA-A scenario for the case of two fiducial models constant in redshift.
The two upper dot-dashed lines refer also to the PCA-A parameterization but with redshift-dependent fiducial models.
The two lower colored solid lines depict the PCA-B scenarios, while the thick solid black line and the blue contours show the mean value and the $1$, $2$ and $3\sigma$ allowed regions within the \pchip\ prescription.}
\end{figure}

Figure~\ref{fig:tau} shows the most relevant results from our analyses of Planck 2015 temperature and polarization data assuming different reionization histories.
As aforementioned, we shall focus on the cumulative redshift distribution function of the reionization optical depth, Eq.~\eqref{eq:cumtau}.
A large departure from $0$ at redshifts $z>10$ would indicate evidence for an early reionization contribution, and therefore for non-standard reionization sources as, for instance, energy injection from dark matter annihilations or from matter accretion on massive primordial black holes.
Notice that the PCA-A method of Ref.~\cite{Heinrich:2016ojb}, in which the PCA decomposition is added linearly to a fiducial $x^{\rm{fid}}_{\rm{e}}(z)$, leads \emph{always} to an early contribution to the optical depth $\tau$, i.e.\ $\tau$ is significantly different from 0 at $z>10$, in contrast to standard reionization scenarios.
Furthermore, the presence of this early contribution is independent of the fiducial model,
as we can see from
the four PCA-A cases depicted in Fig.~\ref{fig:tau}, which provide the same predictions at $z>10$, differing only mildly at small redshifts, regardless whether the fiducial model is a constant function or it depends on the redshift instead. 

\begin{figure}
\centering 
\includegraphics[width=0.85\textwidth]{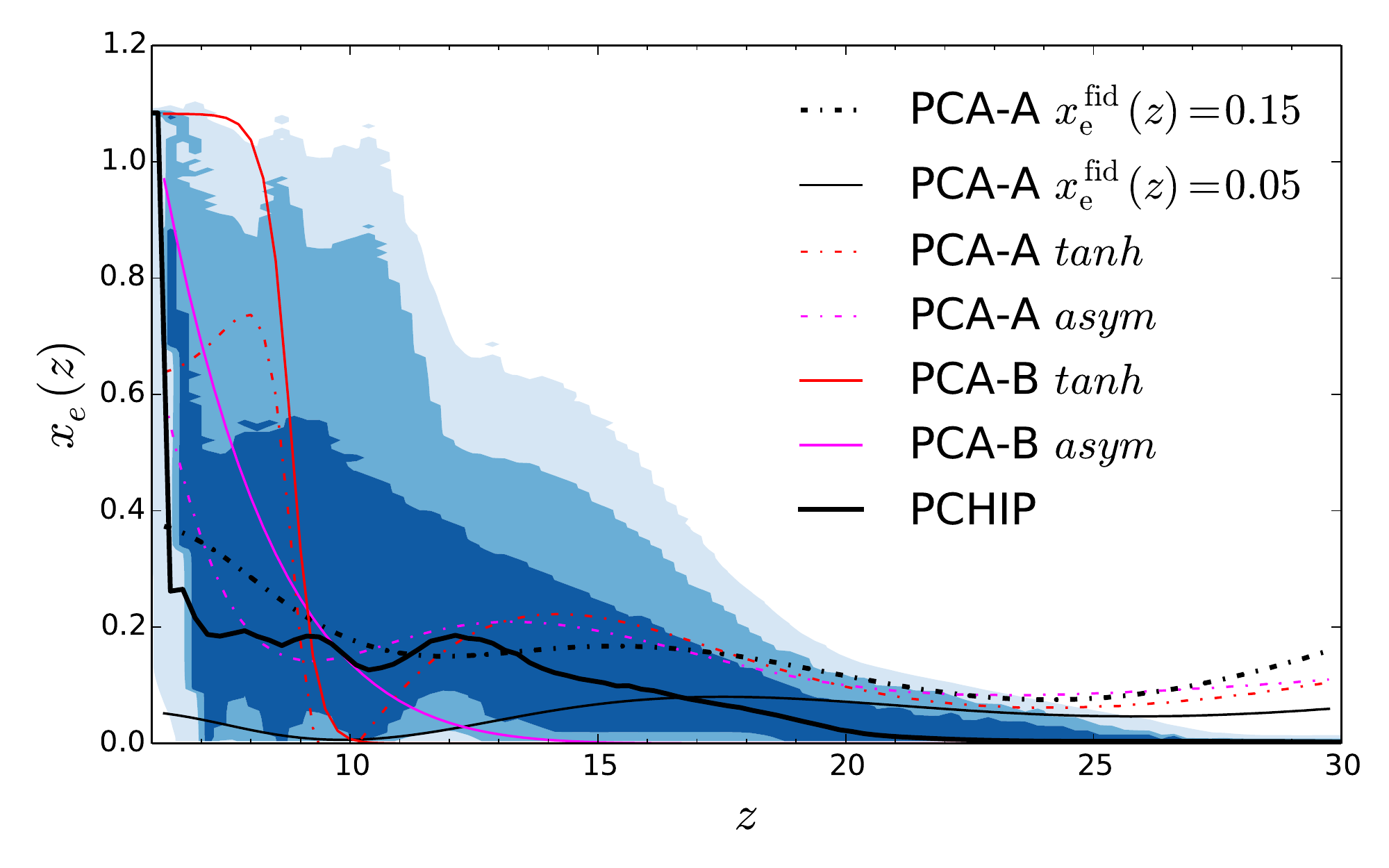}
\caption{\label{fig:xe}
Free electron fraction as a function of the redshift for several possible reionization scenarios.
Line styles and colors are the same as in Fig.\ref{fig:tau}.}
\end{figure}

In order to unravel the origin of this early reionization component present when using the PCA-A description, several tests have been carried out.
Firstly, we have eliminated the physical limits in the PCA amplitudes, finding very similar results.
Secondly, we have simulated mock Planck data with the hyperbolic tangent description and then fitted these data to a PCA-A modeling, using different fiducial models.
We always find two bumps in the recovered $x_e$, see Fig.~\ref{fig:xe}, one located between $z=10$ and $z=15$ and a second one located between $z=20$ and $z=25$.
Upcoming measurements from the Planck satellite could disentangle if this early reionization component is truly indicated by the data or instead it is due to the adopted modeling or to other effects (i.e.\ systematics).

Furthermore, this early reionization component is definitely absent when other possible reionization histories are used in the analyses. 
For instance, in the case of PCA-B parameterizations (see Eqs.~\eqref{eqn:tanh_b} and \eqref{eqn:asym}), there is no early reionization contribution, as $\tau(z)$ is negligibly small for $z>10$.
The same happens for the \pchip\ method, in which the mean reconstructed value of $\tau(z)$ is also very small at high redshifts, showing little evidence for an early reionization component (see also Ref.~\cite{Hazra:2017gtx}).
Notice that the value of $\tau$ today is smaller in the PCA-B approaches than in the PCA-A and \pchip\ descriptions.
However, this behavior is the expected one, as the PCA-B scenarios are very close to those explored by the Planck collaboration in Ref.~\cite{Adam:2016hgk}, where it was found that the current value of $\tau$ is $0.058\pm 0.012$ for the hyperbolic tangent case, in perfect agreement with our findings here, even if we make use of the 2015 Planck likelihood only (the mean value is $\tau=0.068$ for the very same model).
The differences between the PCA-A and PCA-B cases can be understood from the fact that the case B imposes a more restrictive functional form on the ionization history.
 
\begin{figure}
\centering 
\includegraphics[width=0.85\textwidth]{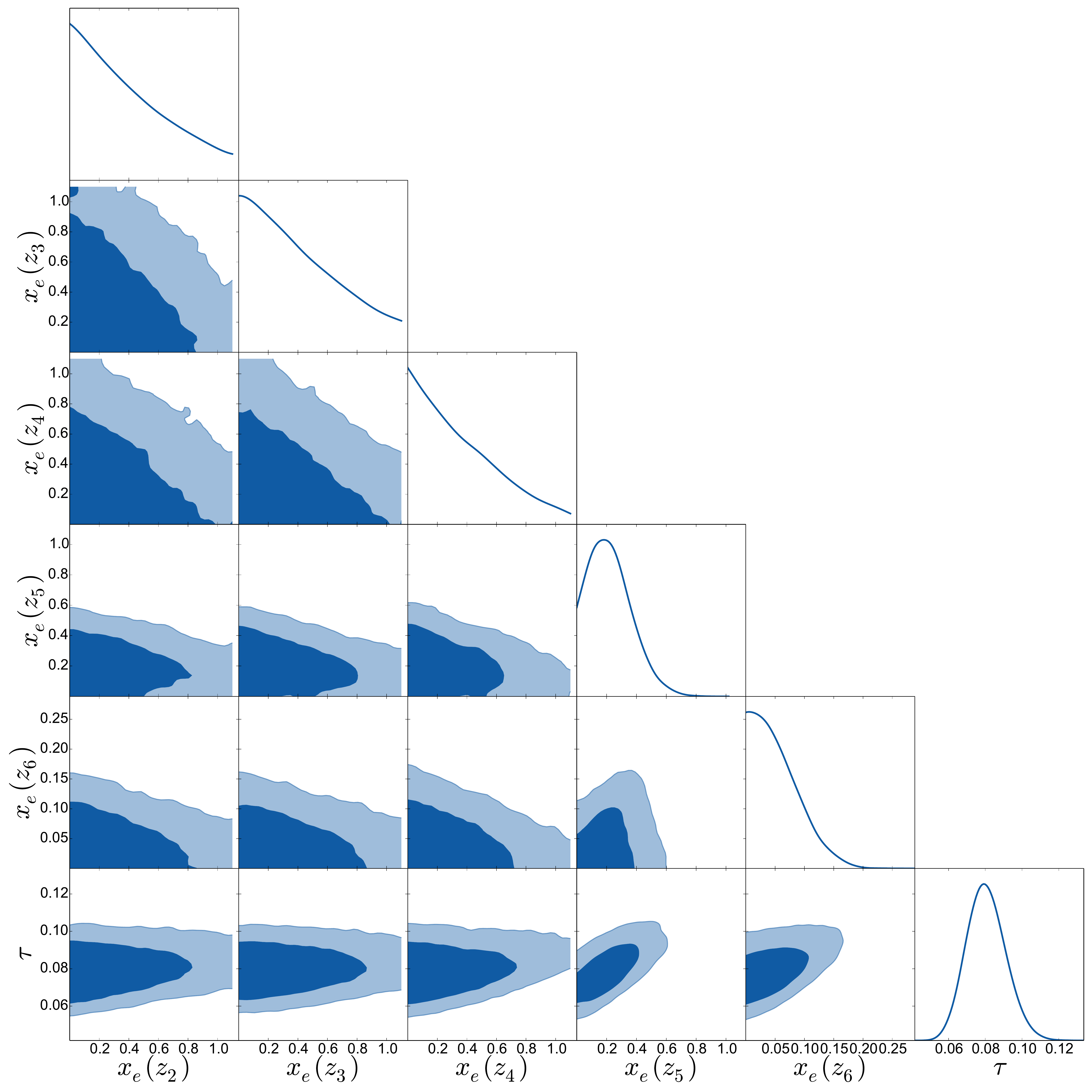}
\caption{\label{fig:pchipamplitudes} $68\%$ and $95\%$~CL allowed regions from the Planck CMB measurements considered here on the amplitudes in the \pchip\  approach, together with the one-dimensional posterior probability distributions.}
\end{figure}

The findings above are fully consistent with our limits on the free electron fraction $x_{e}(z)$ at a given redshift.
Figure~\ref{fig:xe} shows the free electron fraction for the \pchip\ parameterization together with the other PCA-A and PCA-B models explored here.
The color coding is identical to that used in Fig.~\ref{fig:tau}. 
Notice that for the PCA-A models the free electron fraction is almost constant in the redshift interval $z=10-30$, as a consequence of the choice of the fiducial model, and therefore there will always be an early reionization component \emph{within this approach}.
However, when considering either the \pchip\ or the PCA-B models, the free electron fraction is significantly smaller than $0.2$ for redshifts above $z=15$ and it is almost negligible above $z=20$.
Therefore, the fact that current CMB observations need an early contribution to reionization is highly questionable, as it strongly depends on the framework used to analyze the data.
Using Planck CMB temperature and polarization data within the \pchip\ analysis,
we find $x_e<0.90$, $<0.49$ and $<0.13$ at $2\sigma$ in the nodes at $z=10$, $13$ and $20$, respectively.
Fluctuations in the lower $1\sigma$ limits shown in Fig.~\ref{fig:xe}
are numerical artifacts that appear when computing the error bands at intermediate positions between the fixed \pchip\ nodes and cannot be considered as significant.
Figure~\ref{fig:pchipamplitudes} shows the $68\%$ and $95\%$~CL allowed regions for the amplitudes of the \pchip\ nodes, i.e.\ the $x_{\rm e} (z)$ at the redshifts listed in Eq.~\eqref{eq:nodes7},
from the Planck CMB measurements considered here.
A quick inspection of Fig.~\ref{fig:pchipamplitudes} tells us that all the amplitudes are perfectly compatible with a vanishing value.
Only one of them, $m_5$, the node corresponding to $z=13$, shows a very mild departure from $0$. However, this mild departure is far from being a significant effect, as it barely appears at $1\sigma$.
We can therefore conclude that there is no evidence for a high redshift component in $x_{\rm e}(z)$.
Notice also from Fig.~\ref{fig:pchipamplitudes} that, in general, the \pchip\ amplitudes are anti-correlated among themselves.
We also illustrate the derived distribution for the value of the reionization optical depth, $\tau_{\rm PC}$, which is significantly correlated with the nodes at the higher redshifts.
Even a modest increase of $x_{\rm e}$ at $z=13$ or at $z=20$ would imply a significant shift towards larger values of the current reionization optical depth. 

In order to further assess our findings above, we adopt here two information criteria 
which have been widely exploited
in astrophysical and cosmological contexts (see Refs.~\cite{Liddle:2007fy,Trotta:2008qt} for details), namely the frequentist Akaike Information Criterion (AIC)
\begin{equation}
\textrm{AIC}\equiv -2 \ln \mathcal{L }_{\rm{max}} +2k~,
\end{equation}
which establishes that the penalty term between competing models is twice the number of free parameters in the model, $k$; and the Bayesian Information Criterion (BIC)
\begin{equation}
\textrm{BIC}\equiv -2 \ln \mathcal{L }_{\rm{max}} +k\ln N~,
\end{equation}
in which the penalty is proportional to the number of free parameters in the model times the logarithm of the number of data points $N$.
The best model is the one minimizing either the AIC or the BIC criteria. 
Following Ref.~\cite{Liddle:2007fy}, the significance against a given model will be judged based on the Jeffreys' scale, which will characterize a difference $\Delta$AIC (BIC)$>5$ ($>10$) as a strong (decisive) evidence against the cosmological model with higher AIC (BIC) value.

Adopting first the AIC prescription, we shall compare the different models explored here to the standard scenario, in which reionization is described via just only one parameter, $\tau$. 
This \emph{tau-only} cosmological model gives $-2\ln \mathcal{L }_{\rm{max}}=12956.2$~\cite{Ade:2015xua}.
As a comparison, the PCA-A case with constant fiducial model $x_{e}=0.15$ ($0.05$) provides $-2\ln \mathcal{L }_{\rm{max}}=12954.0$ ($12953.2$). 
Notice that both the PCA-A cases have a higher AIC value than the \emph{tau-only} cosmology because of the larger number of parameters.
The values for $\Delta$AIC are $\Delta$AIC $=5.8$ and $5$, respectively, and therefore there is strong evidence against these possible reionization histories.
Also within the PCA-A description, we get $-2\ln \mathcal{L }_{\rm{max}}=12956.5$ ($12958.3$) in the PCA-A \emph{tanh} (\emph{asym}) fiducial approach.
These two models also provide a larger AIC than the \emph{tau-only} scenario, and again, there will be strong (decisive) evidence against the PCA-A \emph{tanh} (\emph{asym}), in favor of the simplest and most economical \emph{tau-only}  reionization paradigm.
In the case of the \pchip\ approach, our results lead to $-2\ln \mathcal{L}_{\rm{max}}=12954.5$, which 
also indicates strong preference for the \emph{tau-only} scheme. We point out that all the reported values of $-2\ln \mathcal{L }_{\rm{max}}$
are taken from the corresponding MCMC chains, and not from a specific minimization algorithm.
For this reason, they may not be extremely precise and they must be considered only as fair estimates of the true values of each $-2\ln \mathcal{L }_{\rm{max}}$, with possible errors of order unity, as estimated from the different parallel MCMC chains.
In the case of the PCA-B parameterizations,
the difference in the minimum $-2\ln \mathcal{L }_{\rm{max}}$ from the different MCMC parallel chains is too large to give even a fair estimate of the true minimum,
and we decide not to claim any evidence against these two descriptions, for the reasons listed above.
However, we expect that these two models are equally good in fitting the CMB data, at a comparable level with respect to the \emph{tau-only} scenario, as their reionization histories are extremely close to the standard cosmological framework, see Fig.~\ref{fig:xe}.
Nevertheless, given the fact that the number of parameters in the PCA-B scheme is larger, the \emph{tau-only} reionization description, with current data, will always be favored over the PCA-B parameterization.

We can also compare the different reionization descriptions among themselves using the BIC approach, as all of them have the same number of free parameters (five in total) and also the same number of data points.
The result of comparing the PCA-A and \pchip\ scenarios among themselves will always give very weak or inconclusive answers, as none of them in particular is preferred over the other possible formulations.

\section{Conclusions}
\label{sec:conclusions}
Unraveling the reionization period, which is still a poorly known period in the evolution of our universe, is one of the most important goals of current and future cosmological probes.
This is a mandatory step, not only towards a complete understanding of star formation and evolution, but also to answer questions such as the nature of the dark matter component~\cite{Barkana:2001gr,Yoshida:2003rm,Somerville:2003sh,Yue:2012na,Pacucci:2013jfa,Mesinger:2013nua,Schultz:2014eia,Dayal:2015vca,Lapi:2015zea,Bose:2016hlz,Bose:2016irl,Corasaniti:2016epp,Menci:2016eui,Lopez-Honorez:2017csg,Villanueva-Domingo:2017lae}, constraining dark matter properties or the abundance of accreting massive primordial black holes~\cite{Pierpaoli:2003rz,Mapelli:2006ej,Natarajan:2008pk,Natarajan:2009bm,Belikov:2009qx,Huetsi:2009ex,Cirelli:2009bb,Kanzaki:2009hf,Natarajan:2010dc,Giesen:2012rp,Diamanti:2013bia,Lopez-Honorez:2013lcm,Lopez-Honorez:2016sur,Poulin:2016nat,Poulin:2015pna,Ricotti:2007au,Horowitz:2016lib,Ali-Haimoud:2016mbv,Blum:2016cjs,Poulin:2017bwe}.
Currently, the most accurate measurement of the reionization period comes from Cosmic Microwave Background data through a redshift-integrated quantity: the reionization optical depth $\tau$.
The latest measurements of the Planck collaboration provide a value of $\tau = 0.055 \pm 0.009$~\cite{Aghanim:2016yuo, Adam:2016hgk}, which shows a very good agreement with observations of Lyman-$\alpha$ emitters at $z\simeq 7$~\cite{Stark:2010qj,Treu:2013ida,Pentericci:2014nia,Schenker:2014tda,Tilvi:2014oia}.
However, this measured value of $\tau$ may correspond to very different reionization histories.

The most commonly exploited model for the time evolution of the free electron fraction, $x_e(z)$, uses a step-like transition, implemented via a hyperbolic tangent~\cite{Lewis:2008wr}.
Recently, there have been several studies in the literature claiming that Planck 2015 data may prefer a high-redshift ($z>15$) component to the reionization optical depth, implying a clear departure from the hyperbolic tangent picture.
Here we consider a number of possible reionization scenarios, some of them previously explored in the literature, such as the Principal Component Analysis (PCA) approach of  Refs.~\cite{Hu:2003gh,Mortonson:2007hq,Mortonson:2007tb,Mortonson:2008rx,Mortonson:2009qv,Mortonson:2009xk,Mitra:2010sr,Heinrich:2016ojb}, or the \pchip\ framework~\cite{Hazra:2017gtx}. 
We find that the claimed need for an early reionization component from present data is highly debatable, as it is only motivated by a particular set of reionization descriptions.
In other possible reionization prescriptions, equally allowed by data, we do not find such a preference.
To assess this, we have applied the frequentist Akaike Information Criterion (AIC), which provides an unbiased model comparison method.
The AIC results show that there is strong evidence from current data against more complicated reionization scenarios, always favoring the minimal  scenario with the symmetric hyperbolic tangent function and described by one single parameter, the reionization optical depth $\tau$. In other words, current Planck CMB analyses are unable to provide more information beyond that based on a single value of the $\tau$.  Upcoming data from the Planck mission will help in further disentangling the reionization history of our universe.

\acknowledgments
This work makes use of the publicly available \texttt{CosmoMC}~\cite{Lewis:2002ah} and \texttt{CAMB}~\cite{Lewis:1999bs} codes and of the Planck data release 2015 Likelihood Code~\cite{Aghanim:2015xee}. OM and PVD would like to thank the Fermilab Theoretical Physics Department for hospitality.
OM and PVD are supported by PROMETEO II/2014/050, by the Spanish Grants SEV-2014-0398 and FPA2014--57816-P of MINECO and by the European Union's Horizon 2020 research and innovation program under the Marie Sk\l odowska-Curie grant agreements No.\ 690575 and 674896.   
The work of SG was supported by the Spanish grants
FPA2014-58183-P,
Multidark CSD2009-00064 and
SEV-2014-0398 (MINECO),
and PROMETEOII/2014/084 (Generalitat Valenciana).

This manuscript has been authored in part by Fermi Research Alliance, LLC under Contract No. DE-AC02-07CH11359 with the U.S. Department of Energy, Office of Science, Office of High Energy Physics. The United States Government retains and the publisher, by accepting the article for publication, acknowledges that the United States Government retains a non-exclusive, paid-up, irrevocable, world-wide license to publish or reproduce the published form of this manuscript, or allow others to do so, for United States Government purposes. This work made extensive use of the NASA Astrophysics Data System and {\tt arXiv.org} preprint server.

\bibliography{bibliography}
\end{document}